\documentclass[conference]{IEEEtran}
\ifCLASSINFOpdf
\else
\fi

\ifCLASSINFOpdf   
   \usepackage[pdftex]{graphicx} 
   \graphicspath{{../pdf/}{../jpeg/}{./image/}}    
   \DeclareGraphicsExtensions{.pdf,.jpeg,.png,.jpg}   
 \else   
   \usepackage[dvips]{graphicx}   
   \graphicspath{{../eps/}}    
\fi
\usepackage{float}
\usepackage{graphicx}

\begin{document}

%
\title{Analysis of Path Loss mitigation through Dynamic Spectrum Access: Software Defined Radio}

\author{\IEEEauthorblockN{Chandan Pradhan, Garimella Rama Murthy}
\IEEEauthorblockA{Signal Processing and Communication Research Center\\
International Institute of Information Technology, Hyderabad, India \\
Email: chandan.pradhan@research.iiit.ac.in, rammurthy@iiit.ac.in}
}


%


\maketitle

\begin{abstract}
In this paper, an analysis is carried out for a method to mitigate the path loss through the dynamic spectrum access (DSA) method. The path loss is a major component which determines the QoS of a wireless link. Its effect is complemented by the presence of obstruction between the transmitter and receiver. The future cellular network (5G) focuses on operating with the millimeter-wave (mmW). In higher frequency, path loss can play a significant role in degrading the link quality due to higher attenuation. In a scenario, where the operating environment is changing dynamically, sudden degradation of operating conditions or arrival of obstruction between transmitter and receiver may result in link failure.  The method analyzed here includes  dynamically allocating spectrum at a lower frequency band for a link suffering from high path loss. For the analysis, a wireless link was set up using Universal Software Radio Peripherals (USRPs). The received power is observed to increase by dynamically changing the operating frequency from 1.9 GHz to 830 MHz.  Finally the utility of software defined radio (SDR) in the RF front end, to combat the path loss in the future cellular networks, is studied.
\end{abstract}


%
\IEEEpeerreviewmaketitle

\section{Introduction}

The path loss (PL) is the reduction in power density of a radio signal as it propagates in the space \cite{one}. The main factors contributing to path loss include propagation losses caused by the natural expansion of the radio wave front in free space, absorption losses caused when the energy of the radio signal is absorbed by the obstacles in the path and diffraction losses when part of the radio wave front is obstructed by an opaque obstacle. The PL adversely affect the QoS of a wireless link resulting in possible link failure and hence combating the path loss is a crucial task for any wireless system.\par 
    In LTE, link adaptation and power control (PC) are two tools used to mitigate the effect of the PL. The link adaptation refers to dynamically allocate the modulation and coding (MCS) for communication. This is controlled by the eNodeB (eNB) and help tackling path loss in downlink. In downlink, eNB usually transmits with the maximum power \cite{two,three}. In the uplink, PC provides the way to handle the path loss. The PC in uplink includes determining PL by user equipment (UE) through the received symbol received power (RSRP) from the eNB. The UE then compensates for the PL by changing its transmit power accordingly (open-loop PC). The transmit power of UE also includes an MCS dependent offset determined by the eNB (closed-loop PC) \cite{two}.\par 
    With the next generation cellular network, 5G, looking to exploit the higher frequency (or mmW), the problem of path loss can prove to be a bottleneck \cite{four}. The operating environment is highly dynamic. It can encounter sudden degradation in the operating conditions or the arrival of obstruction between transmitter and receiver, resulting in link failure, especially at high operating frequency. In this paper, an analysis of a specific scenario is considered where the already established method to tackle PL proves to be insufficient. \par
     The method analyzed here includes using software defined radio (SDR) with multi-band reconfigurable antennas (automated by the field programmable gate array of the SDR \cite{five}) at the RF front end of the communication systems. The SDR system is capable of dynamic spectrum allocation (DSA) for the communication. This is used to dynamically lower the operating frequency of a communication system operating at high frequency, in case of degradation of link quality. In the conventional communication system, it is not possible to lower the frequency on the fly as it will require to completely reconfigure the architecture of the communication system. SDR provides the freedom of dynamically changing the parameters of the components at the RF front end so as to operate in various frequency bands. This is analyzed in the paper by creating a wireless link using universal software radio peripherals (USRP). \par
     The paper is organized into four sections. In section 2, the description of the experimental setup and observations for the model tested at Signal Processing and Communication Research Center (SPCRC) in IIIT Hyderabad is discussed. The deployment of the system in real scenario is analyzed in Section 3. In the end, conclusions are presented in Section 4.\par

\section{EXPERIMENTAL SETUP }

\subsection{Setup Requirements }
Universal Software Radio Peripherals (USRPs) and a 64 bit PC constitute the hardware requirements for the experimental setup. In our SPCRC lab, Ettus USRP N210 was used. The N210 hardware is ideally suited for applications requiring high RF performance and great bandwidth. Its architecture includes a Xilinx® Spartan® 3A-DSP 3400 FPGA, 100 MS/s dual ADC, 400 MS/s dual DAC and Gigabit Ethernet connectivity to stream data to and from host processors. The RF front end of the USRP consists of a daughter board which act as a transceiver and antennas used to receive the RF signals. Here WBX daughtercard and VERT900 antennas were used. The WBX provides 40 MHz of bandwidth capability and is ideal for applications requiring access to a number of different bands within its range : 50 MHz to 2.2 GHz. VERT900 antennas operate at 824-960 MHz and 1710-1990 MHz bands. The PC used in the experiment was Lenovo ThinkCare with 64 bit processor and 2 GB RAM. The software requirements for the setup require GNU Radio which is an open source software. \par 
    \textit{Note: Though the VERT900 antenna is said to operate in 824-960MHz and 1710-1990 MHz bands, yet during the experiment, reliable data transmission was observed using the VERT900 antenna for the complete band of 824-1990MHZ. }

\begin{figure}[H]
\centering
\includegraphics[width=2.5in]{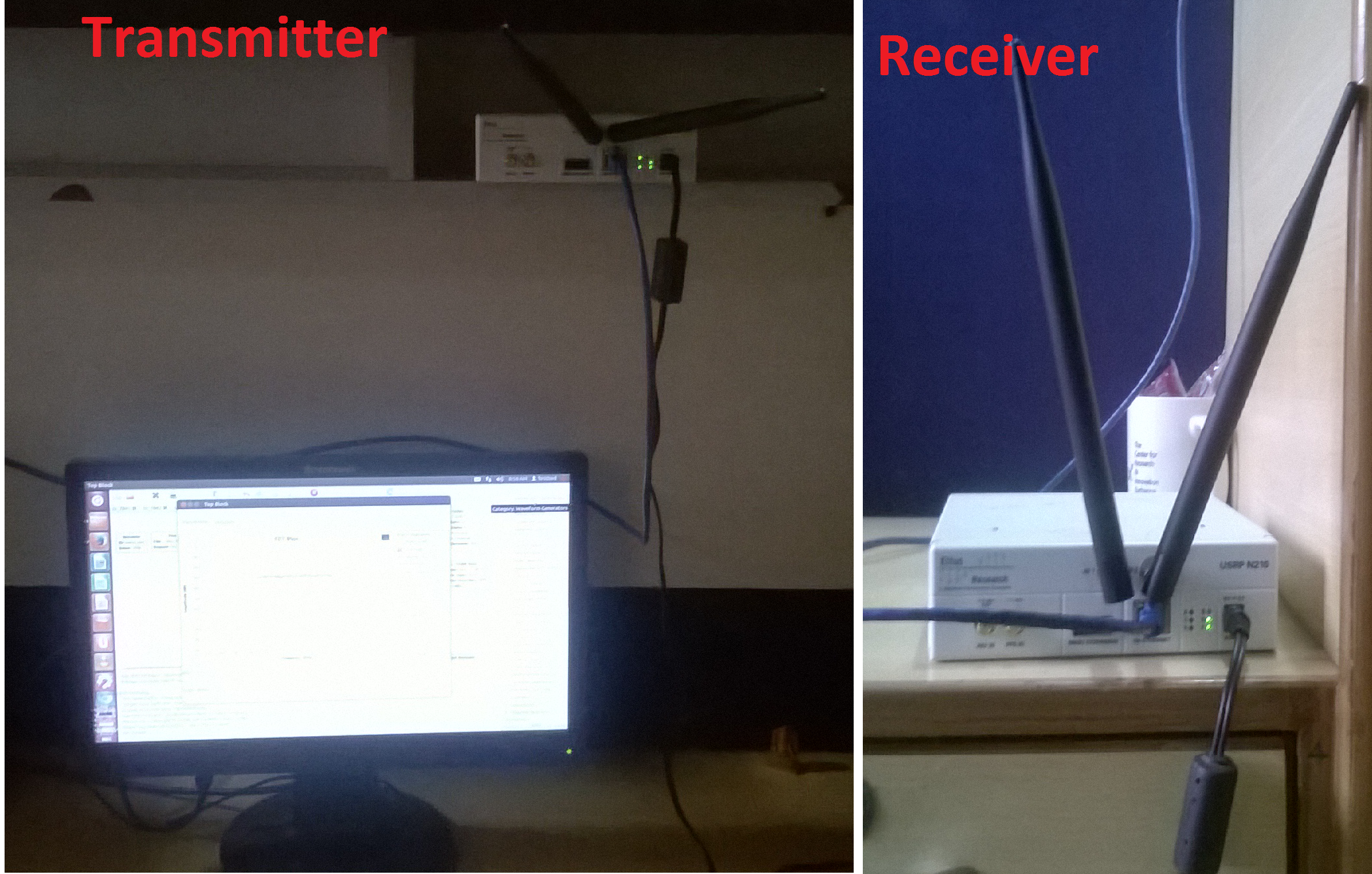}
   \caption{The path loss experimental setup at SPCRC lab in IIIT Hyderabad}
   \label{fone}
\end{figure}

\subsection{Deployment of the Setup}
For the deployment of hardware, as shown in Fig.\ref{fone}, two USRP N210 were connected to the PC through a LAN switch. The LAN switch was connected to the gigabit Ethernet port of the PC. All the connections are made using the Cat 5e cables.  The WBX daughter board was used with the USRPs. VERT900 antennas were connected to the RF1 and RF2 of the USRPs. One of the USRP was used as transmitter and the other as a receiver. Both USRP was at a distance of about 2m from each other. The transmitter was placed 1.5m above the floor and the receiver was placed 1m above the floor. There was no line of sight (LOS) between the transmitter and receiver and they both were separated by two wooden slabs in the lab. The USRPs were interfaced with the PC using the GNU Radio software. The noise floor in the receiver USRP was measured to be -90 dB.\par
    There are two basic experiments conducted to show the utility of SDR for combating the path loss. In the first experiment, the effect of path loss on the received power or received signal strength (RSS) with respect to operating frequency was studied. For the experiment, an unmodulated sinusoidal signal of frequency 1 KHz was transmitted from transmitter to receiver. The RSS was measured for four operating frequencies: 830MHz, 1.2GHz, 1.6GHz and 1.9GHz.  The ITU indoor propagation model was used to analyze the path loss and interpolate values of RSS for other operating frequencies. The ITU indoor model is given by \cite{six}\\
    
\begin{equation}
\label{one}
PL = 20 \log f + N \log d + P_f(n) - 28
\end{equation}                         

where PL is the path loss in decibel (dB), f is the operating frequency in MHz, N is the distance power loss coefficient, d is the distance in meters (m), n is the number of floors between transmitter and receiver and  $P_f (n)$ is the floor loss penetration factor. As the transmitter and receiver are on the same floor, $P_f(n)$ is neglected. Also $PL=P_t-P_r$ , where $P_t$ and $P_r$ are the signal strength at receiver and transmitter respectively. Hence, the equation (\ref {one}) can be written as:\\
 \begin{equation}
\label{two}
P_r = P_t - 20 \log f - N \log d + 28
\end{equation}   
                                
In the experiment, the transmitted power and the distance were kept constant and therefore equation (\ref{two}) can be represented as:\\
\begin{equation}
\label{three}
P_r = \alpha - 20 \log f 
\end{equation}                                                                 
where  $\alpha = P_t -N \log d +28$. Though every factor is kept constant, there can be slight variation in value of the PL with time due to changing environmental conditions which result in minor variation in RSS. Hence equation (\ref{three}) can be represented as a function of time:\\
     \begin{equation}
\label{four}
P_r(t) = \alpha(t) - 20 \log f
\end{equation}                                                   
  where  $\alpha(t) = P_t -N \log d +28+\beta(t)$ and $\beta(t)$ is a time varying factor dependent on operating environment.Four different sets of RSS were measured at time intervals of 30 minutes for four operating frequencies: 830MHz, 1.2GHz, 1.6GHz and 1.9GHz. For each set, $\alpha$ is assumed constant and RSS for the frequency range of 830 MHz to 1.9GHz is calculated through interpolation using measured RSS at the four operating frequencies.  \par
      In the second experiment, the dynamic variation of RSS, of an OFDM system, with changing operating frequency is analyzed. The analysis is carried out according to the LTE standards. For transmission, random bits were generated and were modulated using 16-QAM. The modulated bits were then mapped on to the OFDM symbols. An FFT size of 512 was used for OFDM, out of which 200 tones were occupied. A cyclic prefix of size 128 was used.  The OFDM symbols were then transmitted through the USRP. At receiver the reverse procedure was employed to recover the transmitted bits.  In the experiment, the operating frequency of both the transmitter and receiver was kept as a variable, i.e., it can be varied on-fly between 830MHz and 1.9GHz using the variable slider. Also, the transmitter RF chain gain was set as a variable which can be varied from 0 dB to 40 dB using a slider. The receiver RF chain gain was constant at 10dB.    

\subsection{Observations}
The RSS at frequencies of 830MHz, 1.2GHz, 1.6GHz and 1.9GHz for four sets measured at regular interval of 30 minutes is tabulated in table.\ref{table1}. For each set, in order to compensate for possible variation due to changing environment, the $\alpha_{avg}$  is obtained by taking the average value of $\alpha$ obtained from the four measured RSS using (3).

\begin{table}[H]
\renewcommand{\arraystretch}{1.3}
\caption{Received signal strength (in decibel) for four sets measured}
\label{table1}
\centering
\begin{tabular}{|c||c||c||c||c||c|}
\hline
\bfseries Frequency & \bfseries 830MHz & \bfseries 1.2GHz & \bfseries 1.6GHz & \bfseries 1.9GHz & \bfseries $\alpha_{avg}$ \\
\hline\hline
\textbf{Set1}& -43.09 & -53.53 & -60.85 & -63.15 & 126.59\\
\textbf{Set2} & -43.09 & -55.92 & -61.84 & -62.50 & 126.20\\
\textbf{Set3} & -41.44 & -56.57 & -62.14 & -62.82 & 126.03\\
\textbf{Set4}& -42.70 & -56.57 & -61.84 & -62.10 & 126.23\\
\hline
\end{tabular}
\end{table}

     The plot, using MATLAB, for RSS for the four sets in the frequency range of 830 MHz to 1.9GHz obtained using the interpolation is shown in fig.2 . It can be observed as we go higher in the operating frequency, more is the attenuation suffered by the transmitted signal and hence lesser is the RSS. 
          
     \begin{figure}[H]
\centering
\includegraphics[width=2.5in]{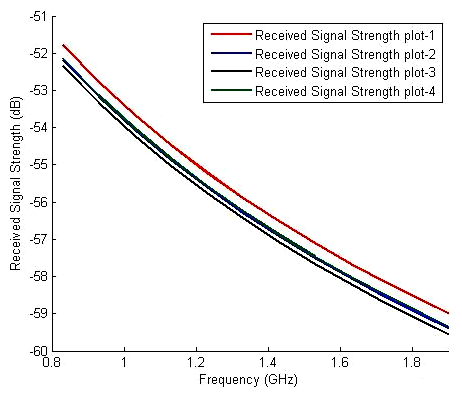}
\begin{center}
    \caption{Received signal strength w.r.t. operating frequency }
\end{center}   
   \label{ttwo}
\end{figure}

\begin{figure}[H]
\centering
\includegraphics[width=2.5in]{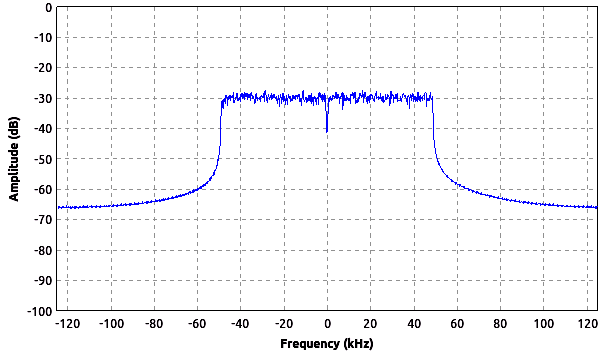}
   \caption{FFT plot for transmitted signal in baseband  }
   \label{fthree}
\end{figure}

\begin{figure*}
\centering
\includegraphics[width=5.25in ,height=2.55in]{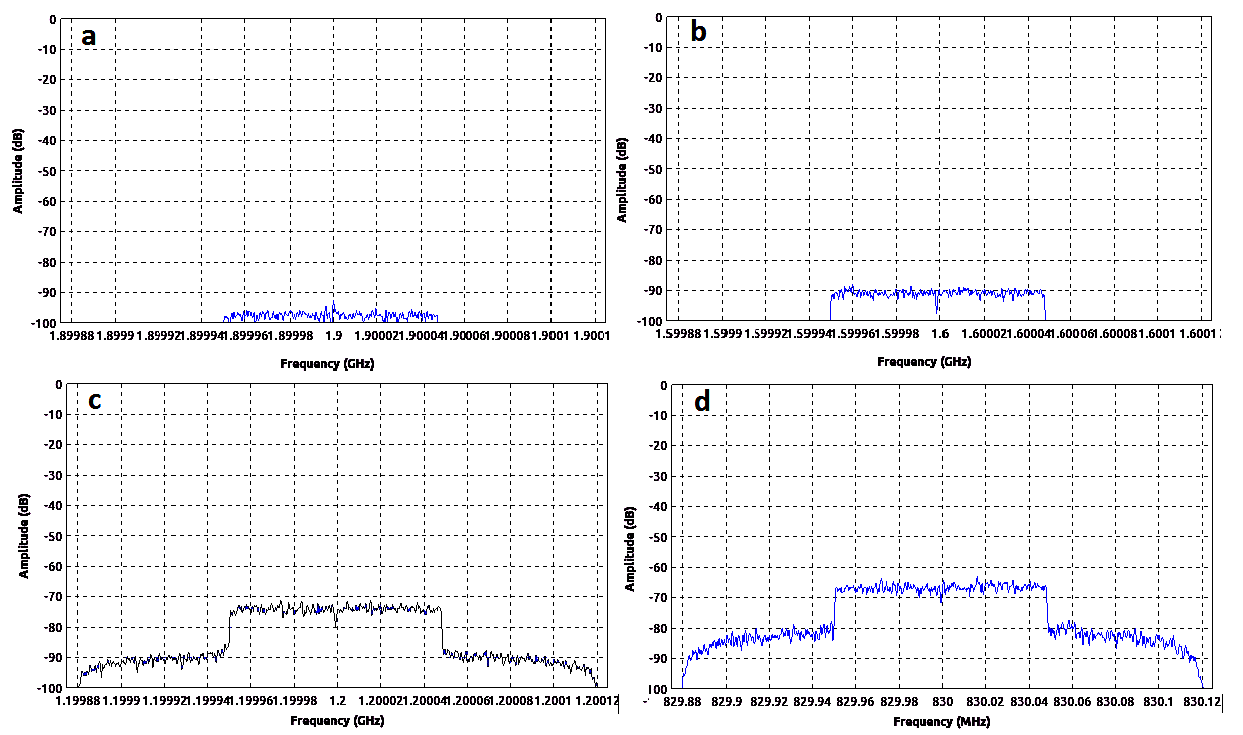}
\caption{FFT plot for received signal strength in pass band for operating frequency of a) 1.9GHz b)1.6GHz c)1.2GHz  d) 830MHz }
\label{ffour}
\end{figure*}

     Next, we look at the second experiment which helps in understanding the utility of SDR in real time application. The FFT of the trasnsmitted signal, in baseband, can be observed in the fig.\ref{fthree}. All the FFT plots henceforth is shown in passband to help observe the plots w.r.t the operating frequency. The transmitter RF chain gain was kept constant at 0dB. The FFT plot of received signal can be seen in fig.\ref{ffour}. The received signal was initially received at 1.9GHz. The transmission signal was then changed from 1.9GHz to 1.6GHz to 1.2GHz and finally to 830MHz. It is observed as the operating frequency is decreased the RSS is increased, resulting in the reliable reception of data. Initially, the RSS at 1.9GHz and 1.6GHz is below the noise floor of -90dB resulting in unreliable reception. At lower operating frequencies, the RSS is above the noise floor.\par
   In the second part of the experiment, the transmitter RF chain gain is varied on-fly. The receiver RF chain gain was kept constant at 10dB. Initially the received signal was received at 1.9GHz and transmitter RF chain gain was at 0dB. The transmitter RF chain gain was then increased on-fly from 0dB to 13dB to 26dB and finally to 40dB. From fig.5, it can be seen that as the gain is increased, the RSS is also increasing. The RSS at 1.9GHz and transmitter RF chain gain of 40dB is comparable to the RSS at 830GHz and transmitter RF gain of 0dB.   
   
\begin{figure*}
\begin{center}
    \includegraphics[width=5.25in, height=1.36in]{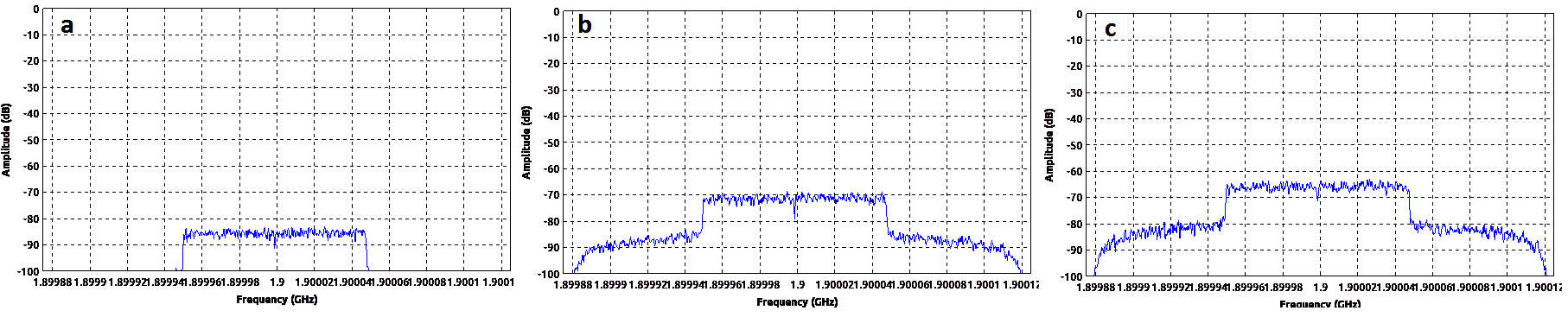}
    \caption{FFT plot for received signal strength in pass band at 1.9GHz for transmitter RF chain  gain  of a) 13dB b)26dB c)40dB  }
    
 \end{center}
\label{ffive}
\end{figure*}

\section{SYSTEM DEPLOYMENT}
\subsection{SDR in Future Cellular Networks}
Extensive research is being done in small cells deploying millimeter-wave (mmW) for communication in future cellular networks. The small cell systems are considered suitable due to low transmit power, short distances and low mobility \cite{four,seven}. The use of large chunks of underutilized spectrum in the mmW bands has gained significant interest in realizing the aforementioned 5G vision and requirements.  Specifically \cite{four} considers the 28- and 38-GHz bands to be initial frequencies where mmW cellular systems could operate. In many dense urban areas, cell sizes are now often less than 100-200 m in radius, possibly within the range of mmW signals based on measurements in \cite{four}.\par
\begin{figure}[H]
\centering
\includegraphics[width=2.5in]{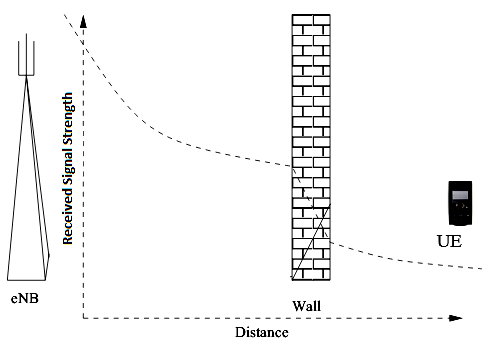}
   \caption{Drop in received signal strength due to the obstruction between eNB and UE}
   \label{fsix}
\end{figure}

     For mmW, path loss plays a significant role in degrading the link quality due to higher attenuation. In a scenario, where in the small cell the operating environment is changing dynamically, sudden degradation of operating conditions or arrival of obstruction between transmitter and receiver may result in link failure (fig.\ref{fsix}). In such a scenario, one solution could be dynamically allocating spectrum resource at a lower operating frequency for a link suffering from high path loss. As observed from the experiment carried out in the previous section, as the operating frequency is decreased, the RSS increases. This is quite helpful when the RSS is below the noise floor due to high attenuation at higher operating frequency. One technology which will complement the deployment of SDR in future cellular network is full-duplex communication \cite{eight,nine}. The full-duplex communication will allow same spectrum resource for both uplink and downlink operation. This allows simultaneous change to the same operating frequency for both uplink and downlink \cite{nine}. \par
    The change in the operating frequency of the communication system is dependent upon metrics which can be decided by the network operators. For example, the decrease in block error rate (BLER) below the desired level or the decrease of RSS below the noise floor can trigger the change in the operating frequency to lower values. Also, as discussed in the previous section, the RF chain gain of both transmitter and receiver can also be dynamically changed to maintain the RSS.\par  
  \textit{ Note: These methods include some challenges and are considered only when current methods of adaptive MCS and PC fails to tackle the problem of path loss at higher operating frequencies.  }
  
\subsection{Challenges in Deployment of the  System}
There are many challenges when deploying SDR in cellular networks. Some of the important challenges include: 1) As the operating frequency is decreased to lower range, there will be increased requirement of bandwidth at lower frequency. Hence it should be ensured that the operating frequency is decreased only when there is availability of enough spectrum resource at lower frequency. 2) The deployment of SDR requires the use of programmable architecture at the eNB and UE front end along with multi-band reconfigurable antennas. This will allow dynamic shift in operating frequqncy. 3) Use of full-duplex in cellular network will increase the computational complexity of the communication system \cite{eight}.
\section{Conclusion}
  In this paper, an analysis was carried out for a method to mitigate path loss through the dynamic spectrum access (DSA) method. The method analyzed included dynamical allocation spectrum at a lower frequency for a link suffering from high path loss. For the analysis, a wireless link was set up using Universal Software Radio Peripherals (USRPs). The RSS is observed to increase by dynamically changing the operating frequency from 1.9 GHz to 830 MHz. Also the RSS was observed to increase for an increase in transmitter RF chain gain. Finally the utility of software defined radio (SDR) in the RF front end, to combat the path loss in the future cellular networks, was studied.





%



\end{document}